\documentclass[pdflatex,sn-mathphys-num,referee]{sn-jnl}
\usepackage{graphicx}%
\usepackage{multirow}%
\usepackage{amsmath,amssymb,amsfonts}%
\usepackage{amsthm}%
\usepackage{mathrsfs}%
\usepackage[title]{appendix}%
\usepackage{xcolor}%
\usepackage{textcomp}%
\usepackage{manyfoot}%
\usepackage{booktabs}%
\usepackage{algorithm}%
\usepackage{algorithmicx}%
\usepackage{algpseudocode}%
\usepackage{listings}%
\usepackage{siunitx}%
\usepackage{braket}%
\newcommand\identity{1\kern-0.25em\text{l}}%
\begin{document}%
\title{Probing Local Topology in a Disordered Higher-Order Topological Insulator}%
\author*[1]{\fnm{Johannes} \sur{Düreth}}\email{johannes.duereth@uni-wuerzburg.de}%
\author[1]{\fnm{Simon} \sur{Widmann}}%
\author[1]{\fnm{Philipp} \sur{Gagel}}%
\author[1]{\fnm{Siddhartha} \sur{Dam}}%
\author[1]{\fnm{Simon} \sur{Betzold}}%
\author[1]{\fnm{Monika} \sur{Emmerling}}%
\author[1]{\fnm{Christian G.} \sur{Mayer}}%
\author[1]{\fnm{David} \sur{Laibacher}}%
\author[1]{\fnm{Martin} \sur{Kamp}}%
\author[2]{\fnm{Oleg A.} \sur{Egorov}}%
\author[2]{\fnm{Ulf} \sur{Peschel}}%
\author[3]{\fnm{Tobias} \sur{Hofmann}}%
\author[3]{\fnm{Ronny} \sur{Thomale}}%
\author[4]{\fnm{Alexander} \sur{Cerjan}}%
\author[1]{\fnm{Sven} \sur{Höfling}}%
\author*[1]{\fnm{Sebastian} \sur{Klembt}}\email{sebastian.klembt@uni-wuerzburg.de}
\affil[1]{\orgdiv{Technische Physik, Wilhelm-Conrad-Röntgen-Research Center for Complex Material Systems and Würzburg-Dresden Cluster of Excellence ct.qmat}, \orgname{University of Würzburg}, \orgaddress{\street{Am Hubland}, \city{Würzburg}, \postcode{97074}, \state{Bayern}, \country{Würzburg}}}%
\affil[2]{\orgdiv{Institute of Condensed Matter Theory and Optics}, \orgname{Friedrich-Schiller-Universität Jena}, \orgaddress{\street{Max-Wien Platz 1}, \city{Jena}, \postcode{07743}, \state{Thüringen}, \country{Germany}}}%
\affil[3]{\orgdiv{Institute for Theoretical Physics and Astrophysics and Würzburg-Dresden Cluster of Excellence ct.qmat}, \orgname{University of Würzburg}, \orgaddress{\street{Am Hubland}, \city{Würzburg}, \postcode{97074}, \state{Bayern}, \country{Würzburg}}}%
\affil[4]{\orgdiv{Center for Integrated Nanotechnologies}, \orgname{Sandia National Laboratories}, \orgaddress{ \city{Albuquerque}, \postcode{87123}, \state{New Mexico}, \country{USA}}}%
\abstract{
Higher-order topology is prized for its ability to realize lower-dimensional boundary states which are stable beyond fine-tuning.
However, disorder presents a failure mechanism that can destroy topological in-gap states.
Here, we investigate a disordered two-dimensional polariton lattice and employ the spectral localizer framework to define a real-space topological index rooted in crystalline spatial symmetries. 
This framework enables direct real-space mapping of topology beyond conventional momentum-space classifications, confirming the presence of corner and edge modes in this generalized Su-Schrieffer-Heeger model.
Furthermore, it can directly quantify topological protection of a state.
We leverage the versatility of our platform to experimentally realize normally distributed, random disorder and find that the corner states persist until the spectral gap closes.
Experimentally, this corresponds to a disorder strength of approximately one quarter of the spectral gap.
The spectral localizer accurately identifies the disorder strength at which the bandgap closes, establishing the framework as a predictive tool for every finite size system.
Our results broaden the design principles for higher-order topological insulators and open the way towards implementing disorder-resilient devices for robust lasing, light-routing, and quantum computation.}%
\maketitle
Topological phases of matter are celebrated for their unconventional boundary phenomena and robustness against disorder \cite{kane2005,bernevig2006,ozawa2019,konig2007}.
These systems support charge and spin transport that remain stable in the presence of non-magnetic impurities, owing to the existence of boundary-localized electronic states protected by topological invariants.
Such protection ensures quantized and resilient conduction properties that distinguish topological materials from their conventional counterparts.
The concepts underpinning topological phases extend far beyond electronic systems.
In photonics, analogous principles have enabled the realization of optical structures that host boundary-confined modes offering promising routes toward unidirectional and disorder-immune light transport \cite{haldane2008,wang2009}. 
These photonic topological materials have opened a new paradigm for manipulating light.

However, porting topology from electronic to photonic platforms presents fundamental challenges.
Most photonic systems cannot access many symmetry classes, such as those that involve particle-hole symmetry, which are essential for defining electronic topological invariants.
Consequently, photonic implementations often rely on crystalline symmetries to generate edge or corner modes \cite{fu2011,hughes2011,wu2015}. 
These schemes have led to promising applications including forced injection locking of lasers \cite{dikopoltsev2021}, or a topological interface for quantum optics \cite{barik2018}.
In realistic devices, however, crystalline symmetries are inevitably broken to some extent by fabrication imperfections or external perturbations.
Then, because the underlying symmetry is broken, the topological protection of a state is not guaranteed anymore \cite{chen2010}. 
Since spatial symmetries are investigated, exact crystalline symmetries can never be achieved in any realistic device.
Although these states are no longer strictly topological, Weyl’s inequality 

\begin{equation*}
    \left|\lambda_i^{(H+\delta H)}-\lambda_i^{(H)}\right|\leq\Vert H+\delta H - H \Vert = \Vert\delta H\Vert\,,
\end{equation*}

\noindent ensures that their eigenenergies shift only, at most, linearly in the presence of disorder \cite{weyl1912,bhatia1997}.
Here, $H+\delta H$ and $H$ are Hermitian operators, while $\Vert\delta H\Vert$ corresponds to the $L_2$ matrix norm, the largest singular value of the matrix $\delta H$.
Weyl's inequality therefore provides a rigorous bound on the strength of a perturbation required to shift an in-gap state into the bulk, i.e., it provides a speed limit on the evolution of eigenvalues in Hermitian systems under a perturbation to a Hamiltonian.
The in-gap state is stable under small perturbations, which is, for many practical purposes, sufficient for the function of a device.
Consequently, while Weyl's inequality guarantees their spectral separation from the extended bulk states in the presence of disorder, the origin of boundary states remains rooted in topology.

To classify this topology and topological protection in photonic systems, the spectral localizer framework has emerged as a real-space-based approach that can natively handle finite, disordered and gapless systems \cite{loring2015,cerjan2024,cerjan2022}.
This framework enables the computation of local topological markers, including those in the Altland-Zirnbauer classes \cite{kitaev2009,ryu2010} and crystalline symmetries \cite{cerjan2024a}.
Furthermore, it provides quantitative measures of topological protection, even in the absence of a bulk spectral gap \cite{cerjan2022}.

In this work, we apply the spectral localizer to determine the topology and degree of topological protection in a polaritonic higher-order topological insulator (HOTI), featuring one-dimensional edge and zero-dimensional corner modes. 
Our platform emulates a two-dimensional generalization of the Su-Schrieffer-Heeger (SSH) model \cite{xie2021}, implemented in a state-of-the-art GaAs-based microcavity platform.
This platform offers fine control over the potential landscape \cite{weisbuch1992,deng2010,schneider2016,gagel2025} and has previously been used to realize polariton topological insulators \cite{st-jean2017,klembt2018}.
By introducing deliberate symmetry breaking and controlled structural disorder, we directly probe the system's robustness and quantify the persistence of higher-order boundary states.
This analysis demonstrates the spectral localizer's capacity to rigorously assess and precisely predict topological protection in realistic, non-ideal photonic environments which will aid the design and characterization of next generation photonic devices. 


\section*{Spectroscopic investigation of the unperturbed lattice}

We consider a two-dimensional generalization of the paradigmatic 1D SSH model \cite{su1979}, where the intra-cell hopping $t_\mathrm{A}$ is weaker than the inter-cell hopping $t_\mathrm{B}$ \cite{liu2017a,benalcazar2019,xie2018}.
This system has been realized in photonic nanocrystal cavities \cite{ota2019,kim2020b}, microwave cavities \cite{xie2019} or polariton lattices \cite{wu2023,bennenhei2024}, while related $C_6$- and $C_3$-symmetric structures have been widely explored \cite{wu2015,barik2018,jihonoh2018,dikopoltsev2021}.

\begin{figure}[t]
  \includegraphics[width=\linewidth]{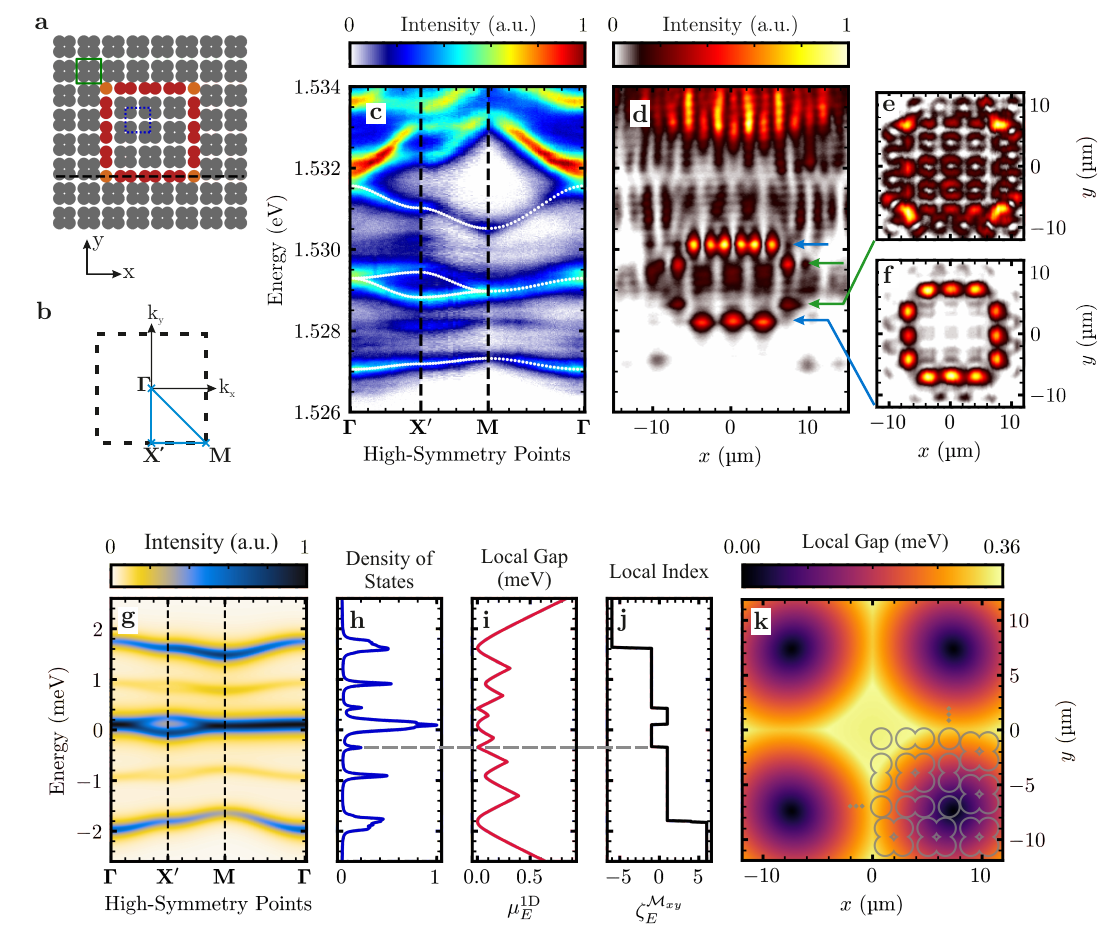}
  \caption{\label{fig:fig1}\textbf{a}, Sketch of a generalized 2D SSH Lattice, featuring a boundary between a compressed bulk unit cell (green solid square) and a topological stretched bulk unit cell (blue dotted square). \textbf{b}, Brillouin zone of the stretched bulk. \textbf{c}, Measurement of the spectrum along the path $\boldsymbol{\Gamma}$-$\mathbf{X'}$-$\mathbf{M}$-$\boldsymbol{\Gamma}$. Dotted lines are calculated Bloch bands. \textbf{d}, Real-space spectrum along the horizontal black dashed line in (\textbf{a}), cutting through the corners (orange) and edge (red). At specific energies, corner states (green arrows) and edge states (blue arrows) are visible. \textbf{e}, and \textbf{f}, Local density of states of the lower-energy corner state and edge state, respectively. \textbf{g-k}, Simulations based on a tight-binding model and the spectral localizer. \textbf{g}, Spectrum based on a tight-binding model along the high symmetry path $\boldsymbol{\Gamma}$-$\mathbf{X'}$-$\mathbf{M}$-$\boldsymbol{\Gamma}$. \textbf{h}, Integrated density of states. \textbf{i}, 1D local gap at the middle of the topological bulk. \textbf{j}, Local index of the $\mathcal{M}_{xy}$ symmetry. It shows a jump of 2 at the energy of the corner, coinciding with a zero in the local gap in (\textbf{i}). \textbf{k}, 2D local gap at $E_\mathrm{corner}$, the inset serves as a reference to the underlying lattice.}

\end{figure}

The lattice (Fig.~\ref{fig:fig1}a) contains a boundary between a stretched unit cell ($t_\mathrm{A}<t_\mathrm{B}$, dotted square), and a compressed one ($t_\mathrm{A}>t_\mathrm{B}$, solid square).
The 2D SSH model undergoes a transition from a trivial ($|t_\mathrm{A}/t_\mathrm{B}|>1$) to a topological phase ($|t_\mathrm{A}/t_\mathrm{B}|<1$) \cite{peterson2020}.
Its topological invariant is the bulk dipole polarization which is quantized due to the presence of reflection symmetries $M_i$, where $i = x,y$ is the mirror axis (c.f. Appendix~\ref{sec:A1}).
Furthermore, the corresponding bulk Hamiltonian  satisfies a $C_4$ rotational symmetry.
Additional diagonal and anti-diagonal reflection symmetries, $M_{xy} = M_xC_4$ and $M_{\bar{y}x} = M_yC_4$, can be constructed from $M_x$ and $M_y$, and $C_4$.
The reflection symmetries $M_i$ used here commute, as opposed to the ones in HOTIs with zero dipole- and nonzero quadrupole polarization.
These require anticommuting reflection symmetries and therefore a phase flux of $\pi$ per unit cell \cite{benalcazar2017}.


Spectroscopy on the polariton 2D SSH lattice reveals corner (orange) and edge (red) states at the interface between topological and trivial regions.
The lattice itself consists of cylindrical potential wells coupled via mode overlap \cite{gagel2025}.
Each potential well has a diameter of $d = \SI{2}{\micro\meter}$.
To accommodate the different inter- and intra-cell hoppings, the wells are spaced by $a_\mathrm{intra} =1.15\cdot d$ and $a_\mathrm{inter} =0.80\cdot d$ for the stretched bulk, resulting in a lattice constant of \SI{3.9}{\micro\meter}.
Within the compressed bulk, $a_\mathrm{intra}$ and $a_\mathrm{inter}$ exchange their respective values.

Using Fourier space spectroscopy (for further information, refer to Section~\ref{sec:Methods}), the band structure of the 2D SSH lattice, including the domain boundary, is measured.
The high-symmetry path $\boldsymbol{\Gamma}$-$\mathbf{X'}$-$\mathbf{M}$-$\boldsymbol{\Gamma}$, along which the measurement is taken, is illustrated in Figure~\ref{fig:fig1}b.
 
In the measured spectrum of the lattice, presented in Figure~\ref{fig:fig1}c, four bulk $s$-bands are observed.
Calculated Bloch bands highlight the bands and appear as dotted lines in Figure~\ref{fig:fig1}c.
Due to the limited spatial extent of the SSH-like domain boundary, its spectral signal appears comparatively weak. 
Nevertheless, additional intensity within the spectral gap of the first and second band is observed around the $\mathbf{X'}$-point.
This feature is attributed to the dispersive edge mode localized at the interface.
To further analyze the sample, a real-space mode tomography is performed.
An energy-resolved spatial cut along the domain boundary (dashed line in Fig.~\ref{fig:fig1}a) is shown in Fig.~\ref{fig:fig1}d, where 0D corner states (green arrows) are clearly visible at the domain corners, lying within the spectral gap between edge (blue arrows) and bulk bands.

The low-energy corner state resides in a spectral gap between the edge mode (cf. Fig.~\ref{fig:fig1}f) and the second bulk band.
Because of its finite linewidth, the mode partially overlaps with neighboring bulk states.
The spectral separation between corner states and bulk states already indicates a breaking of chiral symmetry, facilitated by the interface between compressed and stretched bulk regions, where hopping to the trivial bulk differs from the inter- and intra-cell hoppings.
The real-space distribution of the eigenstates at $E_\mathrm{corner}$ (Fig.~\ref{fig:fig1}e) shows dominant emission from the four corners of the boundary, with minor contributions from adjacent bulk and edge states.
The Supplementary Information S1 supports our findings with numerical calculations within the mean-field model for intracavity photons coupled to quantum-well excitons.
Additionally, we show corner state lasing in the Supplementary Information S2.


While spectroscopic measurements provide access to the dispersion and mode profiles of the lattice, the observed corner- and edge modes are merely products of the underlying topology. 
Therefore, experimental findings, when not able to directly map the topology \cite{guillot2025,guillot2025a}, should always be corroborated by theory.
To do this, we employ the spectral localizer framework, a tool to probe the local topology of a system \cite{cerjan2024}.

Physically, this framework classifies a system's crystalline topology based on which inequivalent atomic limit -- a collection of isolated sites distinguished by the Wyckoff positions of their localized wavefunctions -- the system can locally be continued to without closing a spectral gap or breaking a relevant symmetry. 
With this, it is possible to define an energy-resolved topological index for crystalline phases $\zeta^\mathcal{S}_E$, defined in terms of a mirror symmetry $\mathcal{S}$, calculated from the signature of the spectral localizer.
Changes in $\zeta^\mathcal{S}_E$ are integer-valued and correspond to the number of topological states.
Associated with this comes a local measure of topological protection $\mu$, as by Weyl's inequality it is impossible for a perturbation $ \delta H$ to change the system's topology if $\Vert\delta H\Vert < \mu$.

For the calculations of the spectral localizer, a tight-binding model system based on the 2D SSH lattice in Figure~\ref{fig:fig1}a is used.
The hopping amplitudes are derived from experiments, and the method is outlined in the Supplementary Information S3.
All hoppings with a magnitude greater than 0.05\,\% of the strongest bond ($t_\mathrm{B} = \SI{0.855}{\milli\eV}$) are considered to also take chirality breaking terms into account.

As evidenced by the calculated high symmetry path (Figure~\ref{fig:fig1}g) and the density of states (Figure~\ref{fig:fig1}h), the system exhibits edge states along the interface sites (red) and corner states on the corner sites (orange) of Figure~\ref{fig:fig1}a.
The theoretical model reproduces these measurements.

Using a single position operator in the 1D local gap $\mu_{E}^\mathrm{1D}$ or $\zeta_E^{\mathcal{M}_{xy}}$ forces the respective variable to concentrate on the center of the real-space reflection symmetry $\mathcal{M}_{xy}$.
The 1D local gap $\mu_{E}^\mathrm{1D}$ is plotted in Figure~\ref{fig:fig1}i as a function of energy.
It vanishes at the corner-state energy, a necessary condition for the change of the local index.
The smallest protective gap is towards the second 
bulk band and has a size of $\sim 0.175\,$meV, while the the protective gap to the edge states is larger with $\sim 0.275\,$meV.

The local index corresponding to the mirror symmetry $\mathcal{M}_{xy}$ is shown in Figure~\ref{fig:fig1}j.
By construction, the symmetry-reduced spectral localizer, and hence the local index, focuses on the mirror axis of the system, a diagonal axis through the middle of the stretched bulk for $\mathcal{M}_{xy}$.
At the corner state energy, the local index $\zeta_E^{\mathcal{M}_{xy}}$ changes by two.
This indicates the presence of two unpaired topological eigenstates.
To confirm that this corresponds to two of the corner states, we calculate the 2D local gap of the system (cf. Figure~\ref{fig:fig1}k) at this energy.
The local gap is vanishingly small at the position of each of the four corners, indicating the presence of nearby eigenstates.
Because of the underlying $C_4$ symmetry we conclude that all four corner states are topological.

\section*{Breaking of $C_4$-symmetry}
After confirming the formation of topological corner states at the interface between the stretched and compressed domains, we deliberately alter the lattice geometry to probe how symmetry reduction impacts their existence. 
Importantly, the system's boundary has to preserve the protecting symmetry, while the system must be terminated in a way that does not cut through the unit cell.
Removing the upper right unit cell provides a controlled way to break the $C_4$-symmetry while retaining the $M_{xy}$ symmetry along the diagonal \cite{imhof2018}.
This termination is also compatible with our choice of unit cell.
Therefore, we expect a single topological corner state at the lower left corner.

\begin{figure}[t]
  \includegraphics[width=\linewidth]{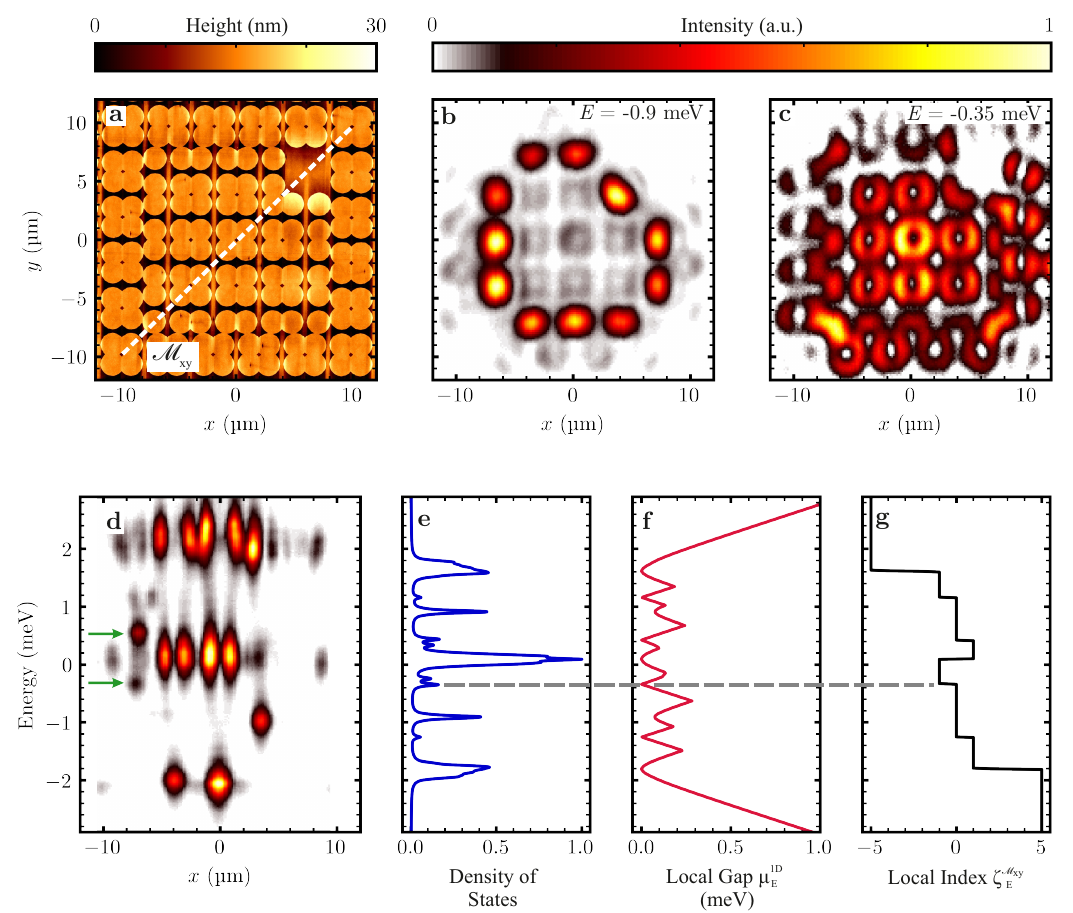}
  \caption{\label{fig:fig3}\textbf{a}, AFM measurement of a polariton lattice with broken, where only the $M_{xy}$ symmetry is retained. \textbf{b}, \textbf{c}, Real-space intensity distributions of the lower energy edge- and corner state, respectively. \textbf{d}, Spectrum along the diagonal, white dashed line in (\textbf{a}), projected onto the $x$-axis. There is only a corner state at the lower left site (green arrow). \textbf{e}, Simulated local density of states. \textbf{f}, Local gap at the position of the lower left corner with a zero crossing at the energy of the corner state. \textbf{g}, Local index for the only remaining symmetry $\mathcal{M}_{xy}$, at the position of the lower left corner site. It changes by one at the energy of the corner state.}

\end{figure}

An atomic force microscope (AFM) measurement of this device is shown in Figure~\ref{fig:fig3}a, where the preserved $M_{xy}$ symmetry is indicated by the dashed white line.
The top right corner of the inner (stretched) lattice is intentionally missing.

When the mode distribution is spectrally resolved, the edge mode adapts to the modified interface by going around the intentional defect (Figure~\ref{fig:fig3}b).
A qualitatively different effect is observed for the corner mode. 
As shown in Figure~\ref{fig:fig3}c, in addition to the unchanged corners in the lower-left, upper-left, and lower-right, no new corner state emerges along the diagonal connecting the lower-left with the missing top-right corner.
This observation is further supported by a spectrally resolved cut along this diagonal (Figure~\ref{fig:fig3}d), projected onto the $x$-axis and shifted by \SI{-1.5263}{\eV} to match the calculations.
In the lower-left corner at $x=$\,\SI{-7}{\micro\meter}, two corner modes are visible: a lower-energy state shown in Figure~\ref{fig:fig3}c and a higher-energy state, indicated by two green arrows.
In contrast, at the obtuse angle of the top right corner no additional corner state is present.
At the new, weakly bound corners, to the right and above of this location, additional corner states arise.
This behavior is expected, since the boundary is terminated consistent with the choice of bulk unit cell.

Tight-binding calculations constructed analogously to those used in Figure~\ref{fig:fig1}g corroborate the experimental observations.
In the cumulative density of states in Figure~\ref{fig:fig3}e, corner states appear below the bulk bands at 0\,meV and edge states at \SI{-0.9}{\milli\eV}.
The energetic splitting of the corner modes arises from differences in the local environments of the newly formed and original corners.
Figure~\ref{fig:fig3}f shows the 1D local gap $\mu_{E}^\mathrm{1D}$ at the lower left corner, calculated as a function of energy.
At the energy of the lower-left corner state, the local gap closes, consistent with a change of the local index $\zeta_E^{\mathcal{M}_{xy}}$ by one, which is plotted in Figure~\ref{fig:fig3}g.
Therefore, the lower left corner supports a topological corner state, even though the $C_4$ symmetry is broken.

\section*{Normally distributed spatial disorder}
Realistic devices might not only feature local defects, like missing unit cells or individual pillars, but also stochastic manufacturing errors that influence positioning accuracy. 
Therefore, the protection of the invariants or the underlying state against disorder is of great interest to the community.
Here, we explore how these states behave under deliberately introduced disorder.
Although studies on this topic exist \cite{xu2006,zhou2020,li2020,agarwala2020,yu2021,hu2021,li2022a,chaou2025}, much of the work focuses on uniformly distributed disorder or configurations that are not directly applicable to realistic devices. 
Instead, we consider normally distributed spatial disorder, i.e., random displacement of lattice sites from their original crystalline position.
The spatial disorder strength $\sigma_\mathrm{s}$ is defined as the standard deviation of the displacement distribution centered around zero.
To quantify the protection of the corner state against disorder, we theoretically investigate 300 disorder configurations and measure at four distinct disorder strengths $\sigma_\mathrm{s} \in \{15,30,60,120\}$\,nm, which is 0.4\,$\%$ to 3.1\,$\%$ of the lattice constant $a = $\,\SI{3.9}{\micro\meter}.


\begin{figure}[t]
  \includegraphics[width=\linewidth]{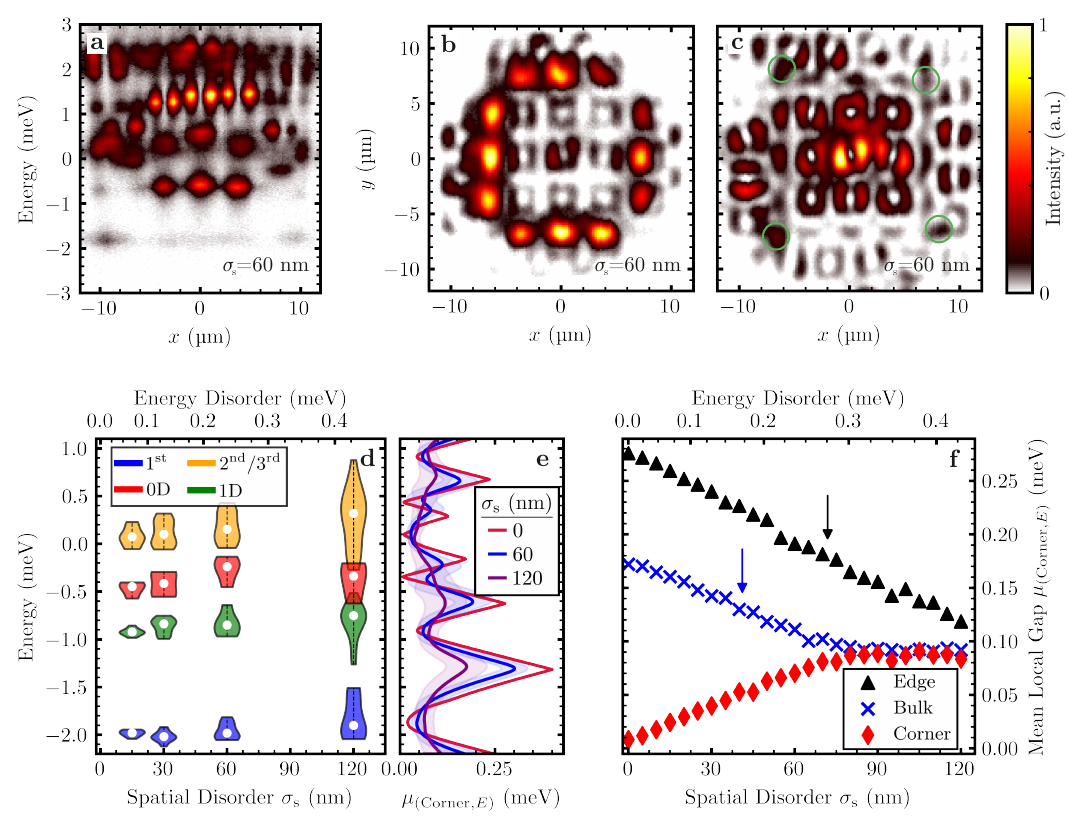}
  \caption{\label{fig:fig4}\textbf{a}, Spectrum of a disordered 2D-SSH lattice along the lower horizontal interface. \textbf{b},\textbf{c}, Iso-energy cut at the energy of the edge- and corner state, respectively. The corner states are marked with green circles for better visibility. Panel (\textbf{a}) to (\textbf{c}) use a disorder of $\sigma_\mathrm{s}=60\,$nm. \textbf{d} ,Analysis of the eigenmode energies at different levels of spatial disorder. For higher disorder strengths, the bulk bandgap closes. \textbf{e} ,Calculated mean local gap at the position of the corner atoms in dependence on energy for 300 different configurations each. Shaded regions are the sample standard deviations. \textbf{f}, Mean local gap (300 configurations) at the position of the corner atoms, at the energy of the corner (red diamonds), at the energy between the corner and edge state (black triangles), and the energy between the corner and second bulk band (blue crosses). Arrows denote the robustness against disorder for the corner (blue arrow) and edge state (black arrow) predicted by the spectral localizer framework.}
\end{figure}

When considering a perturbed lattice, spatial disorder naturally translates into a hopping disorder in the tight-binding model. 
How the hopping disorder is calculated is detailed in the Supplementary Information S4.
In the limit of small perturbations, the eigenenergies of individual sites remain unchanged, while for larger disorder strengths, sites may overlap significantly more and therefore also change their on-site energy.
This development is also indirectly captured by the tight-binding simulations.

A spatial disorder strength of $\sigma_\mathrm{s}=60$\,nm corresponds to an energy disorder of $\sigma_E=0.237$\,meV. 
Figure~\ref{fig:fig4}a shows a horizontal spectral cut through the lower corners of the interface, which closely resembles the unperturbed spectrum shown in Figure~\ref{fig:fig1}d.
It is shifted by \SI{-1.52835}{\eV} to match the calculations.
The main differences are fluctuations in the energetic positions of the corner modes and the non-uniform distribution of the higher-energy edge state.
A cut across the lower edge mode at approximately $-0.9\,$meV, shown in Figure~\ref{fig:fig4}b, reveals that the edge mode remains almost fully intact, besides obtaining a slight inhomogeneity in its spatial intensity distribution.
By contrast, the corner states, displayed in Figure~\ref{fig:fig4}c and highlighted with green circles, show greater apparent variation in space and intensity.
This is due to a comparatively smaller gap to the neighboring bulk states and some variation of the corner states in energy.
Even though the lattice cannot be described by a topological invariant anymore, the corner state is of topological origin and Weyl's inequality ensures its stability against small perturbations.

For a more quantitative assessment of the impact of disorder, the distribution of measured energetic positions of the individual corner-, edge-, and bulk-states at varying spatial disorders $\sigma_\mathrm{s}$ are plotted in Figure~\ref{fig:fig4}d.
Violin plots indicate a spreading of the energies with increasing disorder: At each lattice site, the peak of the energy-dependent intensity measurement is determined. 
The violin plots show the distributions of these fitted energetic peak positions for the first and second $s$-band as well as edge and corner states.
Therein, the white dot marks the mean, and the shaded region shows the distribution of the peaks.
According to these data, energetic disorder is largest for the second and third bulk bands, followed by the first bulk band and edge mode, while the corner state exhibits the smallest variation.
The corner states remain well resolved up to $\sigma_\mathrm{s}=60\,$nm, located in the spectral gap between edge mode and second bulk band, but overlap with both adjacent bands at $\sigma_\mathrm{s}=120\,$nm.
Therefore, the spectral gap between the corner mode and neighboring bands closes between 60\,nm to 120\,nm of spatial disorder, corresponding to an energy disorder of $\sim 0.25\,$meV to $\sim0.43$\,meV.
It should be noted that these measurements represent only one configuration of disorder, and linewidths are not included in Figure~\ref{fig:fig4}d.

Within the spectral localizer framework, the local gap $\mu_{(\boldsymbol{x},E)}$ provides a lower bound on the protection against disorder \cite{cerjan2024a}. 
Specifically, the magnitudes of the maxima of the local gap located spectrally above and below the state of interest (where the local gap is near zero) define the protective gap.
This value gives the minimum norm of the perturbation required to create a state at this position and energy.
Importantly, the local gap can also be calculated in the presence of disorder, even when the energy-resolved crystalline marker is no longer defined.
Figure~\ref{fig:fig4}e shows the 2D mean local gap at a corner site as a function of energy for three different strengths of spatial disorder. 
The shaded regions denote the sample standard deviation of the distribution across 300 disorder configurations.
For zero disorder, the values are about half the respective spectral gap, $\sim0.275\,$meV for the edge mode and $\sim 0.175\,$meV for the second bulk band.
As disorder increases, the local gap at the corner site grows, while the protective gap decreases in value and also shifts in energy.
These dependencies are summarized in Figure~\ref{fig:fig4}f, which shows the mean local gap at the corner energy and position (red diamonds) corresponding to the average energy ``cost" of forming a state at this position and energy. 
An increase of this value points to an increased spectral and local spread of the corner state that naturally occurs for progressively more disordered systems. 
To quantify the protection of the corner state, the maximum local gaps at its respective spectral gaps to the bulk (blue crosses) and edge states (black triangles) is plotted.
Physically, this denotes the required strength of the perturbation to merge either bulk or edge states with the corner state.
Based on the local gap of the unperturbed ($\sigma_\mathrm{s} = 0$\,nm) lattice, the arrows indicate the maximum spatial disorder that the corner states can withstand before merging with either bulk (blue arrow) or edge states (black arrow).
At $\sigma_\mathrm{s} = $ 60\,nm, the energy disorder already exceeds the predicted protection for the corner state but not for the edge mode, in good agreement with the experimental observations.
At higher disorder levels, the corner state merges spectrally with nearby bulk bands and edge states. 
Moreover, hybridization with surrounding states causes the center of mass of the corner mode to deviate from the actual corner site position, further increasing the local gap at that location.
For $\sigma_\mathrm{s}<60\,$nm, disorder effects remain largely local, with the corner state influenced only by its neighbors.
The stronger the disorder, however, the more it is influenced by other states. 
For $\sigma_\mathrm{s}>70\,$nm, the corner state hybridizes with the bulk, and the mean local gaps for bulk and corner states become equal. 
Consequently, the local gap $\mu_{(\boldsymbol{x},E)}$ provides a direct and broadly applicable way to evaluate the effects of disorder in any bounded Hamiltonian.


In this work, we apply the spectral localizer framework to a disordered higher-order topological insulator in an exciton-polariton lattice in order to directly access local topology and identify corner states.
We show that $C_4$-symmetry is not required for the existence of corner states: a single mirror axis intersecting a lattice corner suffices to protect higher-order boundary states.
By intentionally introducing normally distributed spatial disorder, we create and probe a disordered HOTI, observing that corner states persist up to the point where the spectral gap closes.
The threshold of disorder at which corner states disappear is in quantitative agreement with predictions from the spectral localizer framework, and between two- to three-times the hopping strength of the weak hopping amplitude.
These results highlight the robustness of higher-order topological phases under realistic perturbations and demonstrate that the spectral localizer serves as a highly effective diagnostic tool for real photonic platforms. 
Together, these insights open pathways for incorporating disorder-resilient topological modes into photonic architectures, enabling potential applications in robust lasing \cite{ezawa2022}, on-chip light routing \cite{han2024}, and quantum computation \cite{kavokin2022}.

\backmatter

\section*{Methods}\label{sec:Methods}

\subsection*{Sample fabrication}

A sample consisting of 30 mirror pairs of Al\textsubscript{0.15}Ga\textsubscript{0.85}As/AlAs, a $\lambda$-cavity with Al\textsubscript{0.30}Ga\textsubscript{0.70}As, and two stacks of three 13\,nm GaAs quantum wells is grown using a molecular beam epitaxy.
To fabricate the photonic potential, a piece of the sample is spin-coated with a positive polymethyl methacrylate photoresist, exposed using an electron beam, and then developed. 
After this, 20 nm of aluminum is evaporated onto the sample to serve as a hard mask during the etching process. 
The remaining photoresist is removed in a lift-off procedure using pyrrolidone. 
The etchant is a mixture of H$_2$O:H$_2$O$_2$(30\%):H$_2$SO$_4$(96\%)\\ (800:4:1), and the depth is controlled via the etch time.
After etching, the aluminum hard mask is removed in 1\% NaOH. 
To clean from further contaminants, the sample is immersed in 96\% H$_2$SO$_4$ for two minutes. 
Etch depth calibration is performed beforehand using an AFM.
Following this, the sample is placed in a dual ion beam sputtering system (Nordiko 3000) for processing. 
The dielectric top distributed Bragg reflector (DBR), composed of alternating layers of SiO$_2$ and TiO$_2$, is engineered for a central wavelength of $\lambda = 864$\,nm.
For more information on the sample and the fabrication process, please refer to \cite{gagel2025}.

\subsection*{Experimental setup}

Spectroscopic investigations of the 2D-SSH lattices are conducted with the sample held in a liquid helium flow cryostat maintained at $T \approx 5\,$K.
A continuous-wave laser (SolsTiS, M Squared), tuned to the first high-energy Bragg minimum of the stopband, is used for measurements taken both above and below the polariton lasing threshold. 
For momentum-space (k-space) analysis, photoluminescence (PL) is collected using a 20$\times$ objective with a numerical aperture (NA) of 0.4, while real-space imaging is done employing a 50$\times$ objective with an NA of 0.42.

PL is measured in reflection geometry and directed onto the entrance slit of a Czerny–Turner spectrometer (Andor Shamrock SR-750), equipped with an Andor iKon CCD camera.
By adjusting the lens configuration in the detection path, either a real-space image of the sample or a momentum-space image (from the back focal plane of the objective) can be obtained. 
The complete spatial or momentum-resolved PL distribution, I$_\mathrm{PL}(E,x,y)$ or I$_\mathrm{PL}(E,k_x,k_y)$, is recorded by scanning the image across the spectrometer slit using the final lens in the optical setup.

\bmhead{Supplementary information}
Supplementary Information is available.

\bmhead{Acknowledgements}

The Würzburg Team acknowledges financial support by the German Research Foundation (DFG) under Germany’s Excellence Strategy–EXC2147 “ct.qmat” (project id 390858490) as well as DFG project KL3124/3-1 and KL3124/6-1.\\
A.C.\ acknowledges support from the Laboratory Directed Research and Development program at Sandia National Laboratories.
This work was performed in part at the Center for Integrated Nanotechnologies, an Office of Science User Facility operated for the U.S. Department of Energy (DOE) Office of Science.
Sandia National Laboratories is a multimission laboratory managed and operated by National Technology \& Engineering Solutions of Sandia, LLC, a wholly owned subsidiary of Honeywell International, Inc., for the U.S. DOE's National Nuclear Security Administration under Contract No. DE-NA-0003525. 
The views expressed in the article do not necessarily represent the views of the U.S. DOE or the United States Government.

\bmhead{Data Availability}

Experimental data and code used for the analysis are available from the corresponding author upon reasonable request.

\begin{appendices}

\section{A Generalized 2D SSH Model}\label{sec:A1}

The tight-binding Hamiltonian for the topological bulk investigated throughout the manuscript is given by:

\begin{align}
    H(\mathbf{k}) &= \begin{pmatrix}
  0 & q(\mathbf{k}) \\
  q^\dagger(\mathbf{k}) & 0 \\
\end{pmatrix}, \label{eq:h0}\\
q(\mathbf{k}) &=  \begin{pmatrix}
  t_\mathrm{A}e^{ik_x}+t_\mathrm{B}e^{-ik_x} & t_\mathrm{A}e^{ik_y}+t_\mathrm{B}e^{-ik_y} \\
  t_\mathrm{A}e^{-ik_y}+t_\mathrm{B}e^{ik_y} & t_\mathrm{A}e^{-ik_x}+t_\mathrm{B}e^{ik_x} \\
\end{pmatrix} = \begin{pmatrix}
  q_x & q_y \\
  q_y^\dagger & q_x^\dagger \\
\end{pmatrix},
\end{align}

where $\mathbf{k} = (k_x,k_y)$ is the in-plane wave vector and $t_\mathrm{A}$ and $t_\mathrm{B}$ are the intra- and inter-cell hopping amplitudes, respectively.
The lattice constant is set to unity.
The sites in the unit-cell are numbered in the same way as Benalcazar et al. \cite{benalcazar2017}.
The model transitions from a trivial ($|t_\mathrm{A}/t_\mathrm{B}|>1$) to a topological phase ($|t_\mathrm{A}/t_\mathrm{B}|<1$) \cite{liu2017a,peterson2020,xie2019}.
This is accompanied by a change in the bulk dipole polarization from $\mathbf{P} = (0,0)$ to $\mathbf{P} = (\frac{1}{2},\frac{1}{2})$ \cite{liu2017a,benalcazar2019,xie2019}, which is defined as

\begin{align}
P_i = -\frac{1}{(2\pi)^2}\int_\mathrm{BZ}\mathrm{d}^2k \,\,\mathrm{Tr}[\hat{A}_i],\quad i = x,y ,\label{eq:Zak2D}
\end{align}

\noindent where the integration path encloses the Brillouin zone, $\mathrm{Tr}[A]$ denotes the trace of A, $\hat{A}_i(\mathbf{k)}^{mn} = -i\braket{u^{m}_\mathbf{k}|\partial_{k_i}|u^{n}_\mathbf{k}} $ the Berry phase vector potential for $i = x,y$, with $\ket{u^{m}_\mathbf{k}}$ the Bloch eigenvector of band $n$, while $n$ and $m$ are the indices of occupied bands \cite{xie2019}.

From its block off-diagonal form, one can already discern that $H(\mathbf{k})$ obeys chiral symmetry, with $\mathcal{C} = \sigma_3\otimes \sigma_0$ and $\mathcal{C}H(\mathbf{k})\mathcal{C} = -H(\mathbf{k})$.
Here, $\sigma_{1,2,3}$ are the Pauli matrices and $\sigma_0$ the identity matrix.
This chiral symmetry is broken when next-nearest neighbor hopping is considered, or at an interface between a compressed version and a stretched version of the bulk. 
The Hamiltonian also satisfies the following mirror and rotational symmetries:

\begin{align}
    M_xH(k_x,k_y)M_x^{-1} &= H(-k_x,k_y)\\
    M_yH(k_x,k_y)M_y^{-1} &= H(k_x,-k_y)\\
    C_4H(k_x,k_y)C_4^{-1} &= H(k_y,-k_x)\,.
\end{align}

The matrix representations of these symmetries are $M_x = \sigma_1\otimes\sigma_0$, $M_y = \sigma_1\otimes\sigma_1$ and $2C_4 = (i\sigma_2+\sigma_1)\otimes\sigma_0+(\sigma_1-i\sigma_2)\otimes\sigma_1$.
Additional diagonal and anti-diagonal reflection symmetries, $M_{xy} = M_xC_4$ and $M_{\bar{y}x} = M_yC_4$, can be constructed from $M_x$ and $M_y$, and $C_4$.
The reflection symmetries $M_i$ used here commute, as opposed to the ones in multipole HOTIs, which require anticommuting reflection symmetries and therefore a phase flux of $\pi$ per unit cell \cite{benalcazar2017}.

Eigenvalues of Equation~\eqref{eq:h0} can be obtained as the square root of the eigenvalues of $H(\mathbf{k})^2\Psi = E(\mathbf{k})^2\Psi$ as 

\begin{align}\label{eq:eigvals}
    E(\mathbf{k}) &= \pm\sqrt{\delta\pm \sqrt{\alpha^2+\beta^2}},\\
    \delta& = 2t_\mathrm{A}^2-4t_\mathrm{A}t_\mathrm{B}\sin^2{(k_x)}-4t_\mathrm{A}t_\mathrm{B}\sin^2{(k_y)}+4t_\mathrm{A}t_\mathrm{B}+2t_\mathrm{B}^2,\nonumber\\
    \alpha &= 2t_\mathrm{A}^2\cos{(k_x+k_y)}+4t_\mathrm{A}t_\mathrm{B} \cos{(k_x-k_y)}+2t_\mathrm{B}^2\cos{(k_x+k_y)},\nonumber\\
    \beta &= 2(t_\mathrm{A}^2-t_\mathrm{B}^2)\sin{(k_x+k_y)}\,.\nonumber
\end{align}

Equation~\eqref{eq:eigvals} also indicates the chirality of Equation~\eqref{eq:h0}, as the spectrum is symmetric around zero.

A key difference between generalized SSH models and multipole insulators is the spectral position of the corner states.
In a quadrupole insulator, the corner states are located at zero energy by design, while the bulk bands are shifted away from zero energy.
Therefore, the corner states are not obstructed by bulk- or edge-states \cite{benalcazar2017}.
In contrast, in generalized SSH models, the corner states at zero energy can hybridize with bulk bands.
To spectrally isolate them, either next-nearest neighbor hoppings or negative onsite potentials have been introduced \cite{peterson2020,wu2023,bennenhei2024}.
Polariton multipole HOTIs based on the Bernevig-Benalcazar-Hughes model can be designed either by inserting an auxiliary site or by means of orbital hybridization to generate a $\pi$-flux per unit cell \cite{benalcazar2017,keil2016,mazanov2022}. 

\section{The Spectral Localizer}\label{sec:A2}

In the following, qualitative arguments for the working principle of the spectral localizer framework are outlined \cite{cerjan2024}. 
This framework classifies a systems crystalline topology based on which inequivalent atomic limit the system can locally be continued to without closing a spectral gap or violating a relevant symmetry. 
Mathematically, atomic limits manifest as systems whose Hamiltonian $H$ commutes with its position operators $X$ such that $[H^{(\mathrm{AL})} , X^{(\mathrm{AL})}] = 0$, and are characterized in this framework by the signature of the spectral localizer $\mathrm{sig}[L_{\boldsymbol{x},E}]$.

Here, the spectral localizer $L_{(\boldsymbol{x},E)}(\boldsymbol{X}, H)$ is a composite operator made of a system's Hamiltonian $H$ and position operators $\boldsymbol{X} = (X_1,...,X_d)$ in $d$ dimensions constructed from an irreducible Clifford representation $\Gamma_j$ as

\begin{align}
L_{(\boldsymbol{x},E)}(\boldsymbol{X}, H) = \sum_{j=1}^{d}\kappa(X_j-x_j\identity)\otimes\Gamma_j+(H-E\identity)\otimes\Gamma_{d+1},
\end{align}

where the position $\boldsymbol{x}$ and energy $E$ are inputs to the spectral localizer and $\otimes$ denotes the Kronecker product.
The hyperparameter $\kappa$ ensures consistent units between $H$ and $\boldsymbol{X}$ and also weights their contributions to $L_{(\boldsymbol{x},E)}$.
For all calculations $\kappa = \,$0.9/20\,\SI{}{\milli\eV\per\micro\meter}.
For $d=2$, the Pauli matrices serve as the Clifford representation.

Distinct atomic limits correspond to different values of $\mathrm{sig}[L]$, where $\mathrm{sig}[L]$ is the signature of $L$, the number of positive eigenvalues minus the number of negative ones. 
Thus, as the signature of a Hermitian matrix is preserved under perturbations so long as the matrix remains invertible, $\mathrm{sig}[L]$ distinguishes crystalline topology in arbitrary systems for which $[H,X_j] \ne 0 \quad\forall j$, as the system is only homotopy equivalent to an atomic limit with the same $\mathrm{sig}[L]$. 
Therefore, $\mu_{(\boldsymbol{x},E)}$, defined as:

\begin{align}
    \mu_{(\boldsymbol{x},E)}(\boldsymbol{X},H) = \min\left[\left|\mathrm{spec}\left(L_{(\boldsymbol{x},E)}(\boldsymbol{X},H)\right)\right|\right],
\end{align}
where, $\mathrm{spec}(L)$ denotes the spectrum of $L$, is a local measure of topological protection, as by Weyl's inequality it is impossible for a perturbation $\delta H$ to change the system's topology if $\delta H < \mu$. 

The energy-resolved topological index for crystalline phases,
\begin{align}
    \zeta_E^\mathcal{S} = \frac{1}{2} \mathrm{sig}\left[\left(H-E\identity+i\kappa X\right)\mathcal{S}\right],
\end{align}
is defined in terms of a mirror symmetry $\mathcal{S}$ for which $\mathcal{S}^2 = I$, $[H,\mathcal{S}] = 0$ and $\left\{X,\mathcal{S}\right\} = 0$.
Changes in $\zeta^\mathcal{S}_E$ are integer-valued and correspond to the number of unpaired topological states localized at the mirror axis.




\end{appendices}


\bibliography{DisorderedHOTI_Zotero}

\end{document}